\documentclass[11pt,onecolumn,,superscriptaddress,notitlepage]{revtex4-1}

\usepackage{graphics} 
\usepackage{graphicx} 
\usepackage{color} 
\usepackage{amsmath}
\usepackage{amssymb}
\usepackage{amsfonts}
\usepackage{sidecap}

\begin{document} 

\title{On the degeneracy of ordered ground state configurations of the
  aspherical Gaussian core model}

\author{Davide Pini} 
\affiliation{Dipartimento di Fisica ``A.~Pontremoli'', Universit\`a di Milano, Via Celoria 16, 
20133 Milano, Italy}
\author{Markus Wei\ss enhofer} 
\affiliation{Institut f\"ur Theoretische Physik and Center for Computational Materials Science (CMS),
TU Wien, Wiedner Hauptstra\ss e 8-10, A-1040 Wien, Austria} 
\affiliation{Fachbereich Physik, Universit\"at Konstanz, Universit\"atsstra{\ss}e 10, D-78464 Konstanz, Germany}
\author{Gerhard Kahl} 
\affiliation{Institut f\"ur Theoretische Physik and Center for Computational Materials Science (CMS),
TU Wien, Wiedner Hauptstra\ss e 8-10, A-1040 Wien, Austria} 


\begin{abstract}

We provide rigorous evidence that the ordered ground state
configurations of a system of parallel oriented, ellipsoidal
particles, interacting via a Gaussian interaction (termed in
literature as Gaussian core nematics) {\it must} be infinitely
degenerate: we have demonstrated that these configurations originate
from the related ground state configuration of the corresponding
symmetric Gaussian core system via a suitable stretching operation of
this lattice in combination with an arbitrary rotation. These findings
explain related observations in former investigations, which then
remained unexplained. Our conclusions have far reaching consequences
for the search of ground state configurations of other nematic
particles.

\end{abstract}

\date{\today}

\maketitle

\section{Introduction}
\label{sec:introduction}

Reliable identifications of ground state configurations of soft matter
systems -- i.e., to find the energetically most favourable (ordered)
arrangement of particles at vanishing temperature -- is of paramount
relevance to understand the self-assembly strategies of these systems.
In the overwhelming majority of the related investigation it has been
(and often still is) assumed that the particles are spherically
symmetric. This assumption is mostly due to the fact that the
numerical task of identifying ground state configurations is rather
challenging and, from the computational point of view, rather
expensive, as it amounts to find with the help of suitable numerical
techniques in an efficient and reliable manner the optimal particle
arrangement in the space spanned by ``all lattices'' (see, e.g.,
\cite{Wales:1999,Woodley:2008,Antlanger:2016}).

Meanwhile, soft matter particles can be synthesized in essentially
arbitrary shapes, thus moving beyond simple sphericity (for an
overview see, e.g., \cite{Boles:2016}). This remarkable progress in
particle synthesis urges theoreticians to extend the quest for ordered
ground state configurations also to particles that are aspherical in
their shape and/or in their interactions. The computational power of
present-day computers provides the necessary numerical basis for this
challenge.

Among the pioneering contributions, dedicated during the past years to
investigate systematically the ordered ground state configurations for
a simple aspherical system, we focus in this manuscript on a paper
published by Prestipino and Saija (PS) \cite{Prestipino:2007} who
investigated the ground state configurations of ellipsoidal particles,
which interact via a suitably modified Gaussian potential; the aspect
ratio of these particles will henceforward be denoted by $\lambda$. In
an effort to reduce the complexity of the problem the authors made the
simplifying assumption that the directors of all the particles are
oriented in the same direction -- hence the name ``Gaussian core
nematics'' (GCN).  Despite its simplicity, this model combines
characteristic features of realistic soft particles, which have a
pronounced impact on their self-assembly strategies: (i) asphericity,
as witnessed prominently in liquid crystals \cite{deGennes:1995}, and
(ii) mutual penetrability, reflecting the often rather open internal
architectures of typical soft matter macromolecules, as they are
observed, for instance, in dendrimers or polymers (see, e.g.,
\cite{Louis:2000,Bolhuis:2001,Mladek:2007} and citing articles), and
leading in the most extreme case to so-called cluster forming systems
(see, e.g., \cite{Mladek:2006,Mladek:2007a} and citing articles).

The study by PS \cite{Prestipino:2007} and a subsequent contribution
by Nikoubashman and Likos (NL) \cite{Nikoubashman:2010} on the very
same system have revealed important features which apparently have not
been sufficiently noticed within the community, so far: PS reported
about the emergence of ``topologically different degenerate
structures'', whereas NL observed that, for each value of $\lambda$,
the system undergoes a phase transition at a density which is related
to that of the fcc-bcc transition expected in the spherical system
with $\lambda=1$ by a simple scaling law.

Motivated by our own recent investigations of the ordered structures
formed by yet another ultra-soft system \cite{Weissenhofer:2018}
(based on classical density functional theory
\cite{Evans:1979,Evans:1992,Evans:2016}) where we have encountered --
similar as PS \cite{Prestipino:2007} -- unexpected degeneracy issues,
we have re-considered the problem from scratch. Eventually we provide
in this contribution a formally exact evidence for the two following
facts: (i) the energy of the aspherical system can be mapped exactly
on the energy of the related spherically symmetric system; (ii)
further, we rigorously show that an infinite number of ordered ground
state structures {\it must} exist for the aspherical system. These
arguments allow us to fully understand the phenomena observed by PS
and NL, and are closely related to those formerly developed by
Schiller {\it et al.}  \cite{Schiller:2011, Schiller:2012} in the
context of the crystal phases of hard-core ellipsoidal particles.
Somewhat surprisingly, the relevance of their results to those reported
by PS and NL seems not to have been noticed so far.

The impact of our findings can be appreciated in two respects: on one
side, they provide rigorous justifications of the so far inexplicable
observations and embeds them in a broader context; on the other side,
one has to conclude from our findings that the search of ground state
configurations for the GCN (and undoubtedly also for related systems)
does not provide very conclusive results, as one has to end up with
with an infinitely degenerate ground state.

The manuscript is organized as follows: in Section II we present the
model and in Section III we discuss our formalism which provides
rigorous evidence of the above statements; we also dedicate a short
subsection to the related work published by Schiller {\it et al.}
\cite{Schiller:2011, Schiller:2012}. Section IV is dedicated to a
thorough revision and discussion of the data provided by PS and NL, in
view of our formalism. The manuscript is closed with a concluding
section which also addresses the possible future developments of our
findings.

\section{Model}
\label{sec:model}

We consider the {\it aspherical} Gaussian core model (as used in
Refs. \cite{Prestipino:2007, Nikoubashman:2010}), whose functional
form is given by
\begin{equation}
  \Phi({\bf r}) = \varepsilon \varphi \!
  \left[\frac{r}{\sigma(\hat {\bf r}, \lambda)} \right] \, = 
  \varepsilon \exp\! \left[ - \left(
    \frac{r}{\sigma(\hat {\bf r}, \lambda)} \right)^2 \right] ;
\label{pot}
\end{equation}
in this relation $\hat {\bf r} = {\bf r}/| {\bf r}|$, where ${\bf r}$
is the center-to-center vector of two interacting particles,
$\varepsilon$ sets the length scale, and $\lambda$ characterizes the
asphericity of the particle. Since $\Phi(r)$ assumes a finite value at
$r = 0$, this interaction belongs to the family of ultrasoft,
aspherical potentials.

In particular we choose $\sigma(\hat {\bf r}, \lambda)$ to be given by
\begin{equation}
  \sigma(\hat {\bf r}, \lambda) = 
  \frac{\sigma_{0}} {\sqrt{1+(\lambda^{-2}-1)\cos^{2}\vartheta}} \, ,
\label{sigma}
\end{equation}
where $\sigma_0 = \sigma (\hat {\bf r}, \lambda = 1)$ sets the unit
length.  $\vartheta$ is the angle enclosed by $\hat {\bf r}$ and an
arbitrary, but fixed directional unit vector ${\bf u}$, which
determines the orientation of the particles. We stress once more that
this is assumed from the outset to be the same for all particles. If,
without loss of generality, we identify the $z$-axis with the
direction of ${\bf u}$, then $\cos\vartheta=z/r$ and Eq.~(\ref{pot})
can be rewritten as
\begin{equation}
\Phi({\bf r})=\varepsilon\varphi\!\left(\frac{1}{\sigma_{0}}
\sqrt{x^{2}+y^{2}+\left(\frac{z}{\lambda}\right)^{2}}\right)
\, ,  
\label{pot2}
\end{equation}
so that the iso-surfaces of the potential are ellipsoids of revolution
(spheroids) around the $z$-axis with aspect ratio $\lambda$.

Further, we denote the related spherically symmetric potential by
$\Phi_0(r)$, which is obtained from Eq.~(\ref{pot2}) for $\lambda=1$,
i.e.,
\begin{equation}
  \Phi_{0}(r) = \varepsilon \varphi\! \left( \frac{r}{\sigma_0} \right) =
  \varepsilon \exp\! \left[ - \left(
  \frac{r}{\sigma_0} \right)^2 \right] .
\label{pot0}
\end{equation}
Eq.~(\ref{pot2}) can thus be rewritten as
\begin{equation}
\Phi({\bf r}) = \Phi_{0}({\bf r}_{\lambda}) = \Phi_{0}(r_{\lambda}) \, ,
\label{pot3}
\end{equation}
with

\begin{equation}
  {\bf r}_{\lambda} = \left( x, y, \frac{z}{\lambda} \right) ~~~~ {\rm and} ~~~~
  r_{\lambda}=\sqrt{x^{2}+y^{2}+\left(\frac{z}{\lambda}\right)^{2}} .
\label{rlambda}
\end{equation}

Similarly, the Fourier transform of $\Phi({\bf r})$ (denoted by a
tilde) has the form
\begin{equation}
\widetilde{\Phi}({\bf k})=\lambda\sigma_{0}^{3}\varepsilon
\widetilde{\varphi}\!\left(\sigma_{0}
\sqrt{k_{x}^{2}+k_{y}^{2}+(k_{z}\lambda)^{2}}\right) =
\lambda\widetilde{\Phi}_{0}({\bf k}_{\lambda}) =
\lambda\widetilde{\Phi}_{0}(k_{\lambda}) \, , 
\label{phitilde}
\end{equation}
with

\begin{equation}
  {\bf k}_\lambda = (k_{x}, k_{y}, k_{z}\lambda) ~~~ {\rm and} ~~~~
  k_{\lambda}=\sqrt{k_{x}^{2}+k_{y}^{2}+(k_{z}\lambda)^{2}} .
\label{klambda}
\end{equation}
Clearly, the iso-surfaces of $\widetilde{\Phi}({\bf k})$ are also
ellipsoids of revolution around the $z$-axis with aspect ratio
$1/\lambda$.

The system is characterized by its number density $\rho$, throughout
the temperature is set to zero.

\section{The ground state energy}
\label{sec:ground_state_energy}

\subsection{Mapping of the energy between the aspherical and the spherical
potentials}
\label{subsec:mapping}

Consider an ordered configuration of $N$ aspherical Gaussian particles
at $T = 0$ (ground state configuration); its structure is given by a
Bravais lattice with lattice vectors ${\bf a}_1$, ${\bf a}_2$, and
${\bf a}_3$ and cell volume $v = 1 / \rho$; these vectors can be
collected in a matrix ${\bf A}$:

\begin{equation}
  {\bf A} = ( {\bf a}_1, {\bf a}_2, {\bf a}_3) .
\label{matrixa}
\end{equation}
We can then define the related reciprocal lattice with its lattice
vectors ${\bf b}_1$, ${\bf b}_2$, and ${\bf b}_3$; these vectors are
collected in a matrix ${\bf B}$:
\begin{equation}
  {\bf B} = ( {\bf b}_1, {\bf b}_2, {\bf b}_3) .
\label{matrixb}
\end{equation}

At $T = 0$ the free energy reduces to the internal energy $E$ of the
system, given by

\begin{equation}
\frac{E}{N} = \frac{1}{2} \sum_{\bf m\neq{\bf 0}} \Phi({\bf r}_{\bf
  m}) = \frac{1}{2v} \sum_{\bf n}\widetilde{\Phi}({\bf k}_{\bf n}) -
\frac{1}{2}\Phi(0) \, .
\label{lsum}
\end{equation}
The first sum is taken over all lattice positions of the Bravais
lattice, ${\bf r}_{\bf m} = \sum_i m_i {\bf a}_i$ (with $| {\bf
  r}_{\bf m} | \ne 0$) with ${\bf m} = (m_1, m_2, m_3)$ and the $m_i$
being integers, while the second sum is taken over all lattice
position of the reciprocal lattice, ${\bf k}_{\bf n} = \sum_i n_i {\bf
  b}_i$ with ${\bf n} = (n_1, n_2, n_3)$; again the $n_i$ are
integers.

By virtue of Eqs.~(\ref{pot3}) and (\ref{phitilde}) the above
expression can be reformulated as
\begin{equation}
\frac{E}{N} = \frac{1}{2}\sum_{\bf m\neq{\bf 0}}\Phi_{0}(r_{\lambda,{\bf m}})  =
\frac{\lambda}{2v} \sum_{\bf n}\widetilde{\Phi}_{0}(k_{\lambda,{\bf n}}) -
\frac{1}{2}\Phi_{0}(0) , 
\label{eintani}
\end{equation}
using $\Phi(0) = \Phi_0(0)$ -- see Eq. (\ref{pot3}).  The ${\bf
  r}_{\lambda,{\bf m}}$ and ${\bf k}_{\lambda,{\bf n}}$ are specified
in Eqs. (\ref{rlambda}) and (\ref{klambda}), where $x$, $y$, $z$ and
$k_{x}$, $k_{y}$, $k_{z}$ are identified with the components of the
Bravais lattice vectors ${\bf r}_{\bf m}$ and ${\bf k}_{\bf n}$
respectively.

We now observe is that ${\bf r}_{\lambda,{\bf m}}$ and ${\bf
  k}_{\lambda,{\bf n}}$ can also be regarded as vectors of a Bravais
lattice with primitive vectors ${\bf a}^{0}_{i}$ and ${\bf
  b}^{0}_{i}$, $i=1,2,3$:

\begin{equation}
{\bf r}_{\lambda, {\bf m}} = \sum_i m_i {\bf a}^0_i \equiv {\bf r}^{0}_{\bf m}   ~~~~~{\rm and} ~~~~~
{\bf k}_{\lambda, {\bf n}} = \sum_i n_i {\bf b}^0_i \equiv {\bf k}^{0}_{\bf m} \, .
\label{rk}
\end{equation}
The matrices ${\bf A}_0 = ({\bf a}^0_1, {\bf a}^0_2, {\bf a}^0_3)$ and
${\bf B}_0 = ({\bf b}^0_1, {\bf b}^0_2, {\bf b}^0_3)$ obtained by
collecting these vectors are related to the matrices ${\bf A}$ and
${\bf B}$ by:

\begin{equation}
{\bf A}_0 =  
\begin{pmatrix}
a_{1x} & a_{2x} & a_{3x} \\ 
a_{1y} & a_{2y} & a_{3y} \\ 
\lambda^{\!-\!1}a_{1z} & \lambda^{\!-\!1}a_{2z} & 
\lambda^{\!-\!1}a_{3z}  
\end{pmatrix}
= {\bf D}^{-1}_{\lambda} {\bf A}
\label{A0}
\end{equation}

\begin{equation}
{\bf B}_0 =  
\begin{pmatrix}
b_{1x} & b_{2x} & b_{3x} \\ 
b_{1y} & b_{2y} & b_{3y} \\ 
\lambda\, b_{1z} & \lambda\, b_{2z} & \lambda\, b_{3z}   
\end{pmatrix}
= {\bf D}_{\lambda} {\bf B}
\label{B0}
\end{equation}
where we have introduced the matrix ${\bf D}_{\lambda}$, defined as

\begin{equation}
{\bf D}_{\lambda}=
\begin{pmatrix}
1\ & 0\ & 0 \ \\
0\ & 1\ & 0\ \\
0\ & 0\ & \lambda \ 
\end{pmatrix} \, .
\label{diag}
\end{equation}
Obviously, since ${\bf B}$ is the matrix of the reciprocal lattice
vectors of ${\bf A}$, the same relation holds between ${\bf B}^{0}$
and ${\bf A}^{0}$.

Eqs.~(\ref{eintani}) and (\ref{rk}) provide evidence that the
internal energy $E$ of the {\it aspherical} potential with direct
lattice generated by ${\bf A}$ is mapped exactly onto that of an {\it
  spherical} potential with the direct lattice generated by ${\bf
  A}_{0}$. Under the inverse transformation, the (spherical) neighbor
shells of the lattice of the spherical potential are mapped onto
ellipsoids,

\begin{equation}
 x^{2} + y^{2} + \left(\frac{z}{\lambda} \right)^{2} = {\rm const.} ~~~~~
{\rm and} ~~~~~
 k_{x}^{2} + k_{y}^{2} + (k_{z}\lambda)^{2} = {\rm const.}
\end{equation}
for the direct and reciprocal lattices, respectively. Of course, these
ellipsoids are not neighbor shells. However, the number of lattice
points on each of the ellipsoids is the same as that on the neighbor
shells of the spherical case.

We remark that, according to Eq.~(\ref{A0}), the volume $v_{0}$ of the
primitive cell of the spherical potential is related to $v$ by
$v_{0}=v/\lambda$.  Therefore, the density $\rho_{0}$ of the spherical
potential is related to $\rho$ via

\begin{equation}
\rho_0 = \frac{1}{v_0}
= \frac{\lambda}{v} = \lambda \rho .
\label{densities}
\end{equation}

In the light of the above considerations, the ground state of the
aspherical potential can be obtained straightforwardly, provided that
of the spherical potential is known: the matrices ${\bf A}$ and ${\bf
  A}^{0}$ of the lattice vectors are related by Eq.~(\ref{A0}), and
the lattice constant is determined by Eq.~(\ref{densities}). For
instance, if the ground state of the spherical potential at $\rho_0$
is known to be a bcc crystal, then the edge of its conventional cubic
cell must have a length $\ell_{\rm bcc}$ such that
\begin{equation}
\ell_{\rm bcc} = \left( \frac{2}{\rho_{0}} \right)^{1/3} =
\left(\frac{2}{\lambda\rho}\right)^{1/3} \, ,
\label{ellbcc}
\end{equation}
and the energy of such a crystal is equal to that of the ground state
of the aspherical potential at $\rho$. Similarly, if the ground state
of the spherical potential is a fcc or sc crystal, one has
\begin{equation}
\ell_{\rm fcc} = \left( \frac{4}{\rho_{0}} \right)^{1/3}
= \left(\frac{4}{\lambda\rho}\right)^{1/3} \, , 
\label{ellfcc}
\end{equation}
or
\begin{equation}
\ell_{\rm sc} = \left( \frac{1}{\rho_{0}} \right)^{1/3} =
\left(\frac{1}{\lambda\rho}\right)^{1/3} \, . 
\label{ellsc}
\end{equation}

\subsection{Occurrence of degeneracy}
\label{subsec:degeneracy}

A straightforward implication of the above results is that for any
ground state configuration of the aspherical potential at density
$\rho$ there are actually {\em infinitely many} different Bravais
lattices which have the same energy $E$. Let us say that one such
lattice is determined by the primitive vector matrix ${\bf A}$.  Then,
as shown in the former section, ${\bf A}_{0}={\bf
  D}^{-1}_{\lambda}{\bf A}$ (with ${\bf D}_{\lambda}$ given by
Eq.~(\ref{diag})) is a primitive vector matrix of the lattice of the
spherical potential at $\rho_0 = \lambda \rho$.  Now we rotate this
lattice by multiplying ${\bf A}_{0}$ by an arbitrary rotation matrix
${\bf R}$ to obtain ${\bf A}_{0}' ={\bf R}{\bf A}_{0}={\bf R}{\bf
  D}^{-1}_{\lambda}{\bf A}$, where ${\bf R}$ is given, for instance,
by the conventional parametrization in terms of the Euler angles.
Clearly, ${\bf A}_{0}$ and ${\bf A}_{0}'$ correspond to the same kind
of Bravais lattice.  Finally, we go back to the aspherical potential
at $\rho$ by multiplying ${\bf A}_{0}'$ by ${\bf D}_{\lambda}$ to
obtain ${\bf A}'={\bf D}_{\lambda}{\bf A}_{0}'$, i.e.,
\begin{equation}
{\bf A}'={\bf D}_{\lambda}{\bf R}{\bf D}_{\lambda}^{-1}{\bf A}=
\begin{pmatrix}
r_{11} & r_{12} & \lambda^{\!-\!1}r_{13} \\
r_{21} & r_{22} & \lambda^{\!-\!1}r_{23} \\
\lambda\, r_{31} & \lambda\, r_{32} & r_{33} 
\end{pmatrix} 
{\bf A} \, ,
\label{trasdir}
\end{equation} 
where $r_{ij}$, $i,j=1, 2, 3$ are the elements of ${\bf R}$.

By construction, the points of the lattices generated by ${\bf A}$ and
${\bf A}'$ lie on the same ellipsoidal shells, and the number of
points on each shell is the same for both lattices.
Eq.~(\ref{eintani}) then shows that the internal energy is {\em
  exactly} the same.  However, the two lattices will in general be
different: their points, even though lying on the same ellipsoids,
will have different distances from the center of the ellipsoids. Thus
we have an infinite number of lattices having the same energy $E$.

In reciprocal space, one obtains
\begin{equation}
{\bf B}'={\bf D}_{\lambda}^{-1}{\bf R}{\bf D}_{\lambda}{\bf B}=
\begin{pmatrix}
r_{11} & r_{12} & \lambda\, r_{13} \\
r_{21} & r_{22} & \lambda\, r_{23} \\
\lambda^{\!-\!1}r_{31} & \lambda^{\!-\!1}r_{32} & r_{33} 
\end{pmatrix} 
{\bf B} \, ,
\label{tras}
\end{equation}
where ${\bf B}$ and ${\bf B}'$ are, respectively, the original and the
transformed matrix of the primitive vectors of the reciprocal lattice
of the aspherical potential.

\begin{figure}
\includegraphics[width=7cm]{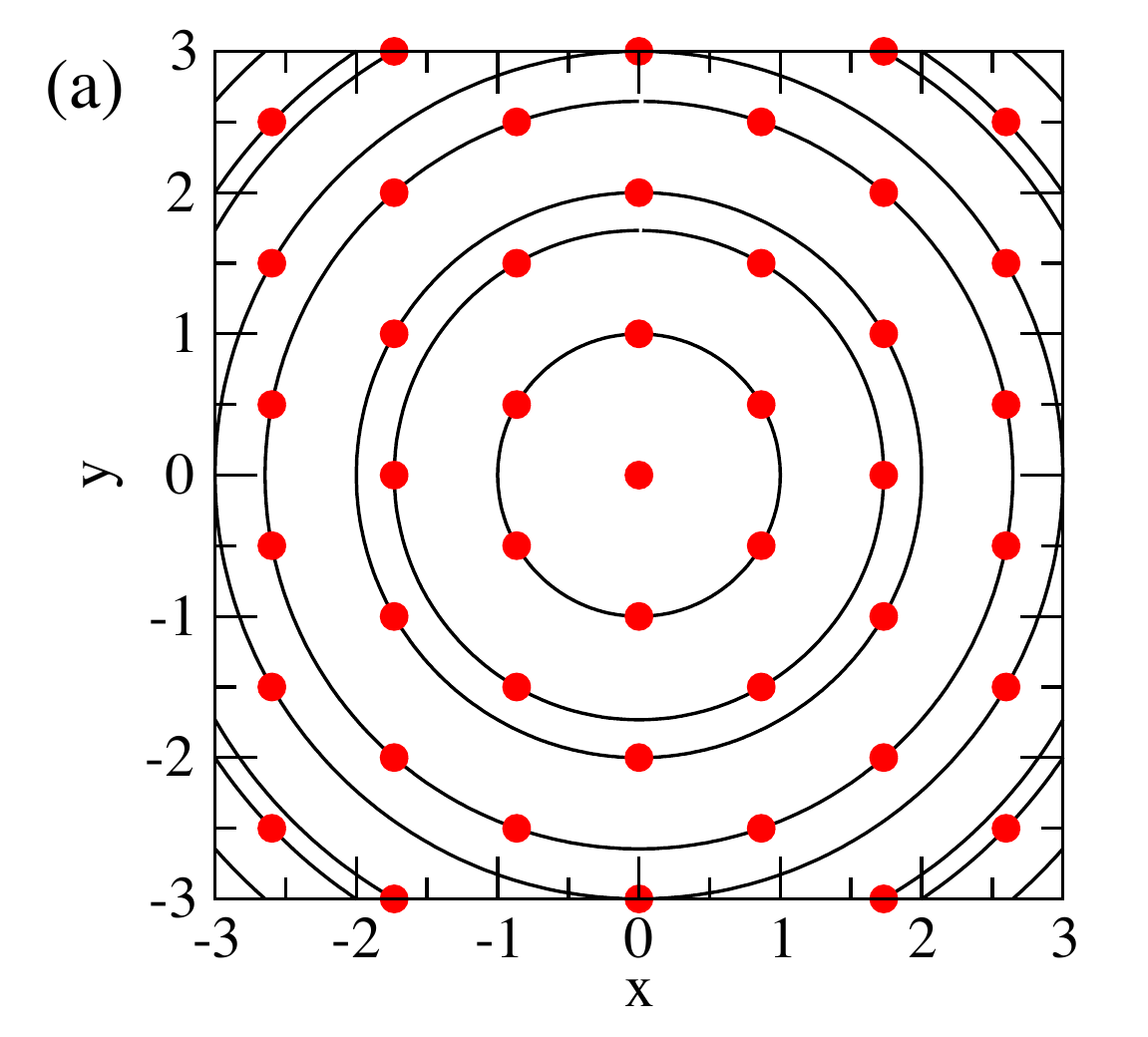}
\includegraphics[width=7cm]{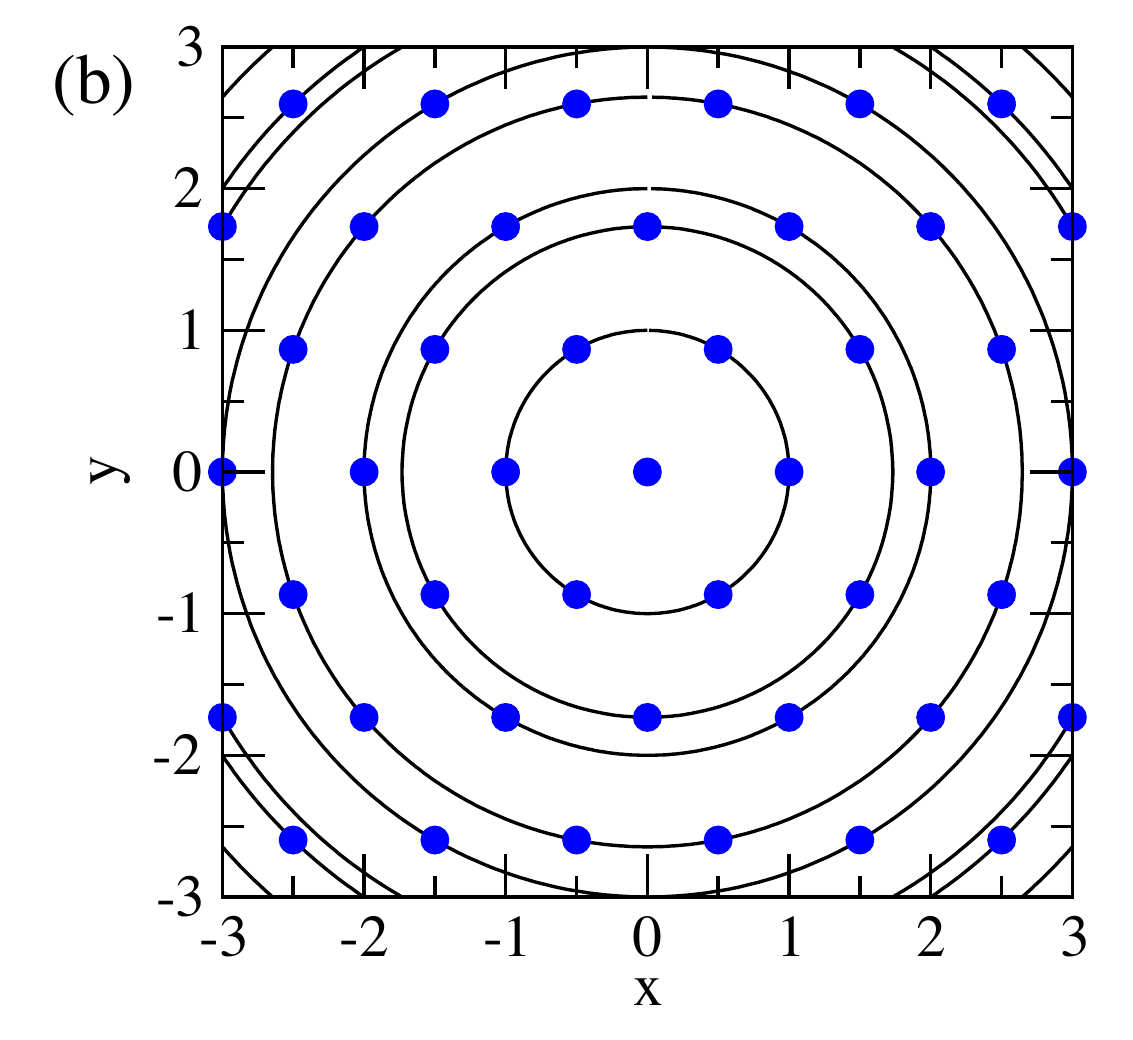}
\includegraphics[width=7cm]{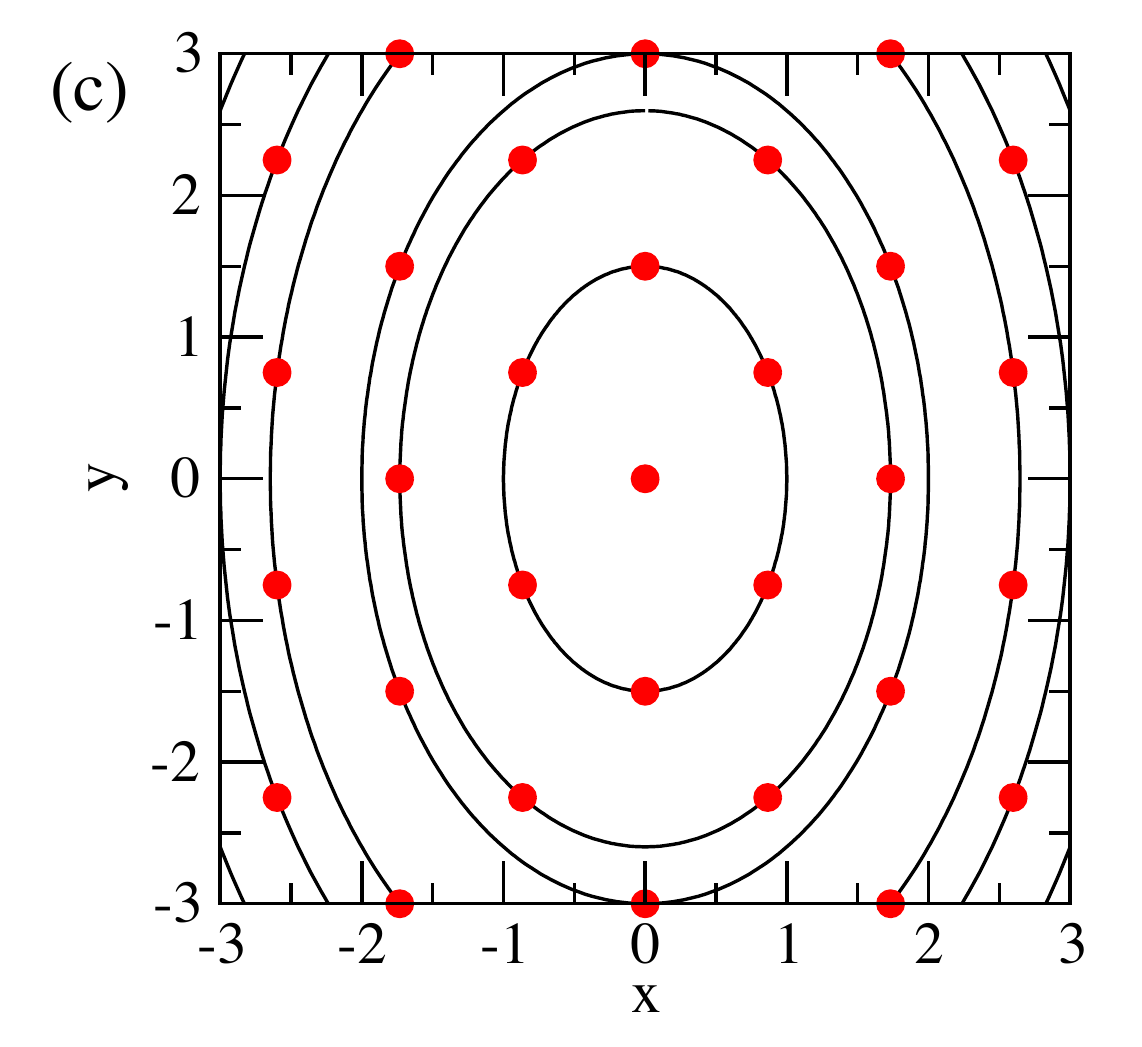}
\includegraphics[width=7cm]{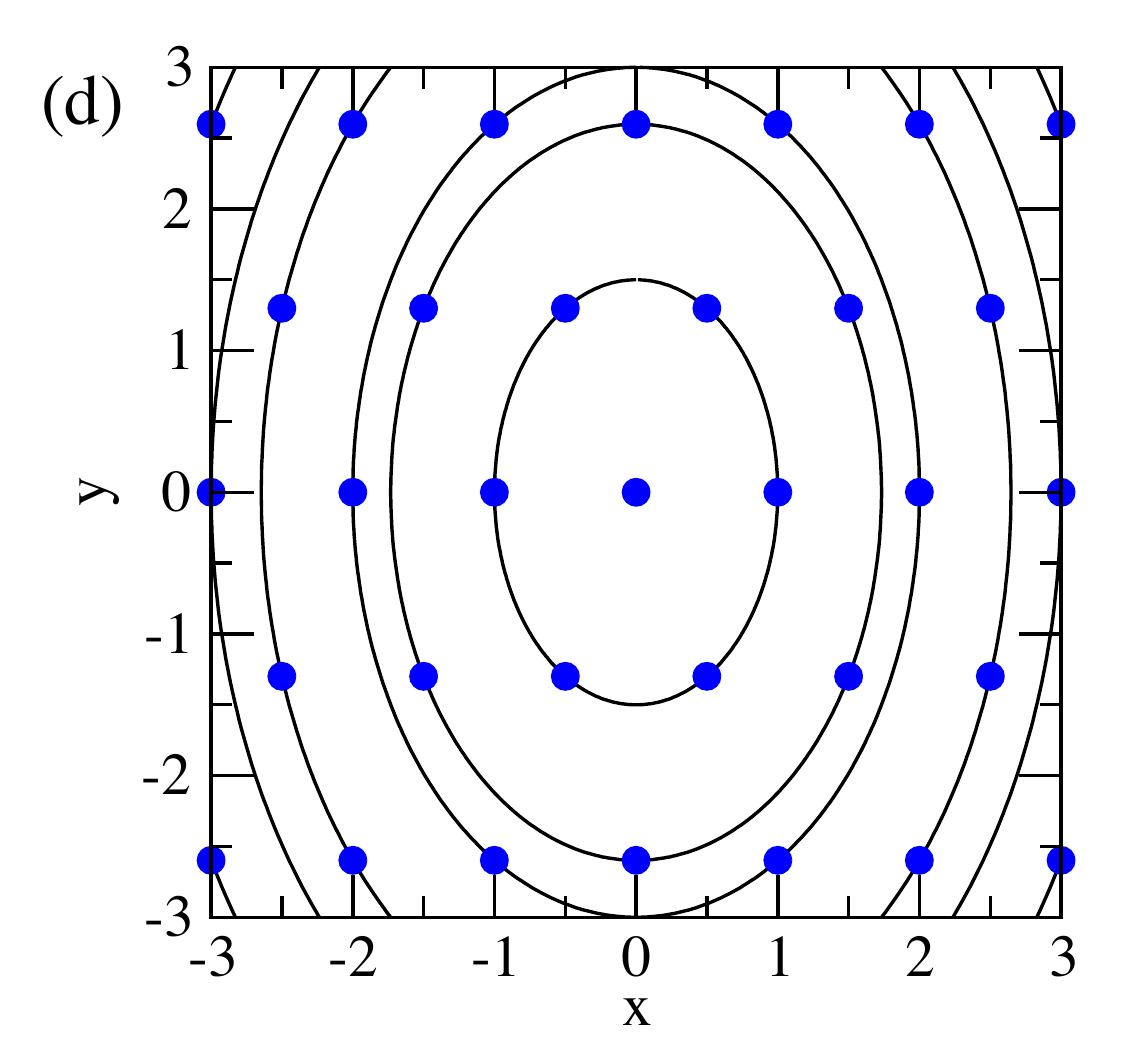}
\caption{The lattices of panels~(c) and ~(d) are obtained by the same
  expansion along $y$ of the triangular lattices shown in panels~(a)
  and~(b) respectively. While panels~(a) and (b) display the same
  Bravais lattice, panels~(c) and~(d) do not.}
\label{fig:2d}
\end{figure}

Fig.~\ref{fig:2d} illustrates what stated above for a two-dimensional
analog: the triangular lattice shown in panel~(a) is rotated by
$90^{\circ}$ to obtain the triangular lattice of panel~(b). The two
lattices actually correspond to the same crystal.  Expanding these
lattices along $y$ by the same factor $\lambda=1.5$ gives the two
lattices shown in panels~(c) and~(d), which have the same number of
points on the same elliptic shells, but do {\em not} describe the same
crystal: for instance, the lattice of panel~(c) has four nearest
neighbors and two next-nearest neighbors, whereas the lattice of
panel~(d) has two nearest neighbors and four next-nearest neighbors;
still the two lattices have the same energy.

In the following, it will be necessary to establish whether, given two
lattices with primitive vectors identified by two matrices ${\bf A}$
and ${\bf A}'$, one can be obtained from the other by the
transformation described above. While this is obviously so if ${\bf
  A}$ and ${\bf A}'$ satisfy Eq.~(\ref{trasdir}), the converse is not
true, i.e., two matrices ${\bf A}$ and ${\bf A}'$ corresponding to
degenerate lattices are not necessarily related by
Eq.~(\ref{trasdir}). The (trivial) reason is that the primitive
vectors of a Bravais lattice are not uniquely determined. We then
proceed to generalize Eq.~(\ref{trasdir}).  To this end, suppose that
the matrices ${\bf C}$ and $\widetilde{\bf C}$ generate the same
lattice. Then for any vector ${\bf m}$ there must be some vector ${\bf
  n}$ such that
\begin{equation}
{\bf C}\cdot{\bf m}=\widetilde{\bf C}\cdot{\bf n} \, ,
\label{equiv}
\end{equation}
and vice versa, where both ${\bf m}$ and ${\bf n}$ are vectors with
integer components. This is equivalent to requiring that both ${\bf
  C}^{-1}\widetilde{\bf C}$ and its inverse must have integer
elements, which in turn is equivalent to requiring that ${\bf
  C}^{-1}\widetilde{\bf C}$ must have integer elements and its
determinant satisfies the relation
\begin{equation}
  \det({\bf C}^{-1}\widetilde{\bf C}) =
  \frac{1}{\det({\bf C}^{-1}\widetilde{\bf C})} \, , 
\label{det}
\end{equation}
i.e., $\det({\bf C}^{-1}\widetilde{\bf C})=\pm 1$. Hence, ${\bf C}$
and $\widetilde{\bf C}$ are related by
\begin{equation}
\widetilde{\bf C}={\bf C}{\bf C}^{-1}\widetilde{\bf C}={\bf C}{\bf L} \, ,
\label{equiv2}
\end{equation}
where ${\bf L}={\bf C}^{-1}\widetilde{\bf C}$ is a matrix with integer
elements such that $\det{\bf L}=\pm 1$. Therefore, Eq.~(\ref{trasdir})
can be generalized to
\begin{equation}
{\bf A}'={\bf D}_{\lambda}{\bf R}{\bf D}_{\lambda}^{-1}{\bf A}{\bf L} \, ,
\label{trasdir2}
\end{equation}
that is
\begin{equation}
{\bf L}=({\bf D}_{\lambda}{\bf R}{\bf D}_{\lambda}^{-1}{\bf A})^{-1}{\bf A}'  
\label{trasdir3}
\end{equation}
for some rotation matrix ${\bf R}$. Eq.~(\ref{trasdir3}) does not
determine this rotation univocally because, for given ${\bf A}$, ${\bf
  A}'$ and $\lambda$, it contains two unknown matrices, namely, ${\bf
  R}$ itself and ${\bf L}$. However, whether it holds or not can be
easily established numerically by spanning the whole group of rotation
matrices, and checking if for some of them the matrix ${\bf L}$
obtained by Eq.~(\ref{trasdir3}) satisfies the above-mentioned
properties.  Actually, in our case one requires from the outset that
the density of the lattices generated by ${\bf A}$ and ${\bf A}'$ be
the same, so one has that $|\det{\bf A}|=|\det{\bf A}'|$.
Eq.~(\ref{trasdir3}) then shows that the condition $\det{\bf
  L}=\pm 1$ is already satisfied, and one has to check only that the
elements of ${\bf L}$ are integer.  Clearly, the matrices ${\bf A}'$
given by Eq.~(\ref{trasdir2}) for the same $\lambda$, ${\bf R}$, ${\bf
  A}$ and different ${\bf L}$ generate the same lattice.

As a final remark, we observe that the considerations put forth in the
present Section can be straightforwardly generalized to aspherical
potentials $\Psi({\bf r})$ with ellipsoidal iso-surfaces of the form
\begin{equation}
\Psi({\bf r})=\varepsilon\psi\!\left(\frac{1}{\sigma_{0}}
\sqrt{\left(\frac{x}{\nu}\right)^{2}+\left(\frac{y}{\mu}\right)^{2}+\left(\frac{z}{\lambda}\right)^{2}}
\right) \, ,
\label{psi}
\end{equation}   
so that the mapping into a spherical potential and the ensuing
degeneracy of the ground state holds also for generic ellipsoids.

\subsection{Relation with previous investigations: Schiller \textbf{\textit{et al.}}}
\label{subsec:schiller}

The findings of Sections~\ref{subsec:mapping}
and~\ref{subsec:degeneracy} are closely related to the results
described in Refs.~\cite{Schiller:2011} and~\cite{Schiller:2012} by
Schiller, Kr\"uger, Wahab, and M\"ogel (SKWM). While the present work
is focused on soft spheroids, those papers were concerned with the
crystal phases of hard-core spheroids with additional attractive or
repulsive interactions, e.g. of dispersion, depletion or electrostatic
origin, which were evaluated in the Derjaguin approximation. When the
size of the ellipsoids is much larger than the range of the forces,
the interaction can be represented as a purely contact potential which
depends on the Gaussian curvature of the ellipsoids at the point of
contact~\cite{Schiller:2011}, so that each particle interacts only
with its first neighbors. SKWM studied the case in which the crystal
consists of parallel ellipsoids, similar to the situation considered
here, and found that ``there exist families of lattices with
geometrically different arrangement of the spheroids but the same
lattice energy''~\cite{Schiller:2012}, including, but not limited to,
close-packed configurations.

The source of this degeneracy was clearly identified in
Ref.~\cite{Schiller:2012}, and is the same as the
rescaling-rotation-rescaling procedure described in
Sec.~\ref{subsec:degeneracy}: by a suitable rescaling of the axes, a
crystal of parallel ellipsoids is mapped into a crystal of spherical
particles.  In doing so, the number density is changed according to
Eq.~(\ref{densities}), but the packing fraction is conserved, since
the volume of the particles and the volume of the primitive cell are
rescaled by the same factor.  Subsequently, the crystal is rotated and
the rescaling is unfolded so as to end up with another crystal of
ellipsoids, see Fig.~2 of Ref.~\cite{Schiller:2012} and
Fig.~\ref{fig:2d} of the present paper.  The structures generated in
such a way can be classified according to the associated lattice of
spherical particles into which the original lattice is mapped.

However, the Gaussian model of this paper differs in two important
respects from the system studied in
Refs.~\cite{Schiller:2011,Schiller:2012}: on the one hand, in the
present case the hard-core interaction between ellipsoids is replaced
by a soft repulsion, whereby each particle does not interact only with
its neighbors, but with all the other particles of the lattice. On the
other hand, the potential considered by SKWM does not depend only on
the center-to-center vector ${\bf r}$ because of the additional
contact forces, which are absent in our case (indeed, they would make
little physical sense for mutually penetrable particles).

As a consequence, the internal energy per particle which they obtain
by mapping the spheroids into spheres by means of the
rescaling~(\ref{A0}) is not necessarily invariant under rotation,
inasmuch as it depends on the contact points of a sphere with its
neighbors via the sum of the squares of their abscissae with respect
to a system with its origin at the center of the sphere, see Eq.~(9)
of Ref.~\cite{Schiller:2012}.  Whether such a quantity is conserved by
a rotation depends on the lattice in which the spheres are arranged:
SKWM find that rotational invariance is fulfilled by lattices which
have at least a threefold, fourfold, or sixfold rotation axis, whereas
it is not fulfilled by lattices with a lower degree of symmetry, such
as the orthormobic, monoclinic, or triclinic lattices.  Lattices of
spheroids whose associated lattice of spherical particles belongs to
the higher-symmetry class have by construction the same energy, even
though their geometrial properties are generally different: in this
case the rescaling-rotation-rescaling procedure does lead to
infinitely many degenerate structures. Instead, for associated
lattices of the low-symmetry class, a rotation leads to a change in
the internal energy, so that the lattices of spheroids obtained by
applying different rotations to the same associated lattice are not
degenerate. Moreover, the degeneracy is removed also in the special
case of associated lattices which have only one threefold, fourfold or
sixfold rotation axis, if the axis ${\bf u}$ of the spheroids is
aligned along it.  In this situation, the rotations around ${\bf u}$
are immaterial as they leave the lattice unchanged, whereas all the
other rotations lead to a change in the internal energy.
 
In summary, for the crystals of hard, parallel spheroids studied by
SKWM establishing whether the ground state is degenerate requires some
knowledge of its structure, specifically concerning the associated
lattice of spherical particles into which it is mapped via the
rescaling procedure.  If the ground state corresponds to a
configuration with a cubic associated lattice, the degeneracy is
certainly present, but other instances are possible, in which the
degeneracy would not occur.

In this respect, the system considered here is simpler, since the
rescaling~(\ref{A0}) maps the potential $\Phi({\bf r})$ into the
spherically symmetric potential $\Phi_{0}(r)$, so that the internal
energy per particle given by Eq.~(\ref{eintani}) is obviously
invariant under rotation for any lattice.  Therefore, the degeneracy
is always present, whatever the lattice. The ground state of the
aspherical Gaussian potential to be discussed in the next Section is
obtained by deforming highly symmetric cubic structures, but the
degeneracy would have been there irrespective of that, unlike in the
case considered by SKWM.

\section{The aspherical Gaussian potential}
\label{sec:discussion_results}

\subsection{General remarks}

On the basis of the considerations put forth above, the phase behavior
of the aspherical Gaussian potential at $T=0$ can be straightforwardly
determined.  As $\tilde \Phi_0(k)$ is a Gaussian and thus positive for
all $k$-vectors, this also holds for $\tilde \Phi ( {\bf k})$ by
virtue of Eq. (\ref{phitilde}). Consequently, the potential at hand
belongs, according to the classification of Ref. \cite{Likos:2001}, to
the class of Q$^+$ potentials: these systems crystallize in
single-occupancy crystals, i.e., their occupancy number $n$ is unity.

The phase diagram of the {\it spherical} Gaussian model has been
studied in detail in
Ref. \cite{Stillinger:1976,Stillinger:1979,Stillinger:1997,Lang:2000,Prestipino:2005,Prestipino:2005a}.
In the following, we will adopt standard reduced quantities by
measuring lengths in units of $\sigma_{0}$ and energies in units of
$\epsilon$. These will be denoted by an asterisk, such as
$T^{*}=k_{\rm B}T/\epsilon$, $\rho^{*}=\rho\sigma_{0}^{3}$, etc.  The
system shows a (reentrant) transition at low temperatures $T$: below a
threshold temperature, the system freezes with increasing density
$\rho_{0}$ either into an fcc lattice and then via a first-order
transition into a bcc lattice, or directly into a bcc; upon further
increasing the density the system melts again. Both types of lattices
show single occupancy ($n = 1$), i.e., no cluster formation can be
observed. Above the threshold temperature (i.e., above $T^{*}\simeq
0.01$) the system never freezes.  At $T=0$, the system forms a crystal
at all densities, and the boundary between the fcc and bcc phases has
been determined analytically~\cite{Stillinger:1979}, the fcc being
favored for $\rho_{0}^{*}<1/\pi^{3/2}$, and the bcc for
$\rho_{0}^{*}>1/\pi^{3/2}$.

The ground state of the aspherical potential is then obtained by
identifying the matrix ${\bf A}_{0}$ of the primitive vectors of the
spherical potential with that of the bcc or fcc lattices ${\bf
  A}_{0,{\rm bcc}}$ or ${\bf A}_{0,{\rm fcc}}$, and deforming ${\bf
  A}_{0}$ according to Eq.~(\ref{A0}). If one adopts the standard
forms for ${\bf A}_{0,{\rm bcc}}$ and ${\bf A}_{0,{\rm fcc}}$, the
expressions of the related matrices ${\bf A}_{{\rm bcc},\lambda}$ and
${\bf A}_{{\rm fcc},\lambda}$ thus obtained are
\begin{equation}
{\bf A}_{{\rm bcc},\lambda}=\frac{\ell_{\rm bcc}}{2}
\begin{pmatrix}
\ 1 & \ 1 &  \! -1 \\
\ 1 & \!\! -1 & \ \ 1 \\
\! -\lambda & \ \lambda & \ \lambda 
\end{pmatrix}   
\, ,
\label{bccani}
\end{equation}
and
\begin{equation} 
{\bf A}_{{\rm fcc},\lambda}=\frac{\ell_{\rm fcc}}{2}
\begin{pmatrix}
  1 & \ 1 & \ 0 \\
  1 & \ 0 & \ 1 \\
  0 & \ \lambda & \ \lambda 
\end{pmatrix}
\, ,
\label{fccani}
\end{equation}
where $\ell_{\rm bcc}$ and $\ell_{\rm fcc}$ are given by
Eqs.~(\ref{ellbcc}) and (\ref{ellfcc}) respectively.  If the above
condition for the occurrence of the bcc or fcc lattices is expressed
in terms of the density $\rho$ of the aspherical model via
Eq.~(\ref{densities}), one obtains that the lattices generated by
${\bf A}_{{\rm bcc},\lambda}$ and ${\bf A}_{{\rm fcc},\lambda}$ are
favored for $\rho^{*}>\rho_{c}^{*}(\lambda)$ and
$\rho^{*}<\rho_{c}^{*}(\lambda)$ respectively, where the reduced
transition density $\rho_{c}^{*}(\lambda)$ is given by
\begin{equation}
\rho_{c}^{*}(\lambda)=\frac{1}{\pi^{3/2}\lambda} \, .
\label{rhox}
\end{equation}
Besides those deformed bcc and fcc lattices, on both sides of
$\rho_{c}^{*}(\lambda)$ there are infinitely many degenerate lattices,
which are obtained by applying to ${\bf A}_{{\rm bcc},\lambda}$ and
${\bf A}_{{\rm fcc},\lambda}$ the transformation~(\ref{trasdir})
involving an arbitrary rotation matrix ${\bf R}$.

In order to ascertain the validity of this picture, we also minimized
the internal energy per particle given by Eq.~(\ref{lsum}) with
respect to the lattice vectors by an independent procedure, i.e.,
without relying on the mapping on the spherical potential described in
Sec.~\ref{sec:ground_state_energy}. The minimization was carried out
numerically by a conjugate-gradient algorithm similar to that used in
Ref.~\cite{Pini:2015} for several values of $\lambda$ and $\rho$,
starting from different initial trial configurations.  We indeed found
that: (i) on convergence of the minimization run, the internal energy
per particle is identical to that of the lattices identified by ${\bf
  A}_{{\rm bcc},\lambda}$ and ${\bf A}_{{\rm fcc},\lambda}$ for
$\rho^{*}>\rho_{c}^{*}(\lambda)$ and $\rho^{*}<\rho_{c}^{*}(\lambda)$,
respectively; (ii) depending on the initial configuration used to start
the minimization, many different degenerate lattices are obtained;
(iii) the matrices ${\bf A}'$ of the primitive vectors of these
lattices are related to ${\bf A}_{{\rm bcc},\lambda}$ and ${\bf
  A}_{{\rm fcc},\lambda}$ by the transformation~(\ref{trasdir2}),
i.e., some rotation matrix ${\bf R}$ could always be identified, such
that the matrix ${\bf L}$ given by Eq.~(\ref{trasdir3}) has integer
elements.

This completes our description of the ground state of the aspherical
Gaussian potential.  In the following, we will discuss in the light of
the present analysis the findings of the previous studies of the same
system performed by PS~\cite{Prestipino:2007} and
NL~\cite{Nikoubashman:2010}.

\subsection{Relation with previous investigations: Prestipino and Saija}
\label{subsec:prestipino}

\subsubsection{Ground state}
\label{subsubsec:PS_ground}

The above conclusions can now be related to the results presented by
PS in Ref.~\cite{Prestipino:2007}, where the aspherical Gaussian model
considered here was first introduced. We point out that according to
PS the nematic direction ${\bf u}$ is identified with the $z$-axis as
here, and that in their notation the asphericity parameter $\lambda$
is equal to $L/D$, $D$ and $L$ being the transverse and longitudinal
diameters of the particles.

PS surmised that the best candidates for the crystal phases of the
model are obtained by deforming the cell of the most common cubic and
hexagonal lattices along a given direction by a factor related to
$\lambda$. This procedure is clearly described by NL in
Ref.~\cite{Nikoubashman:2010} for a cubic lattice as {\em ``The
  stretching of the corresponding [i.e., bcc, fcc, {\em etc.}] cubic
  unit cell along some direction of high crystallographic symmetry by
  a factor $\lambda$ in a volume-preserving fashion, and the
  orientation of the nematics along the stretched axis''}.  According
to PS notation, the vector specifying the direction of stretching is
identified by its coordinates along the axes of the conventional cell,
and the stretching factor is denoted by $\alpha$. So, for instance,
bcc001($\alpha$) is the lattice obtained by deforming by a factor
$\alpha$ the bcc lattice along one of the edges of the cubic cell,
bcc110($\alpha$) that obtained by deforming it along the diagonal of
one of the faces of the cell, and so on.

We observe that in the 001($\alpha$) structure, the direction of
deformation coincides with that of the nematic axis ${\bf u}$, so no
rotation to align the two directions is needed. Therefore, the
001($\alpha$) lattices coincide with those generated by the matrix
${\bf A}$ in Eq.~(\ref{A0}), provided $\alpha$ is identified with the
asphericity parameter $\lambda$, and the non-stretched lattices with
those generated by ${\bf A}_{0}$. In particular, the structures ${\bf
  A}_{{\rm bcc},\lambda}$ and ${\bf A}_{{\rm fcc},\lambda}$, which
according to the previous analysis yield the ground state of the model
at high and low density respectively, correspond in PS notation to the
bcc001($\lambda$) and the fcc001($\lambda$) lattices.

This agrees with the findings of PS. Specifically, they selected a
number of different deformed cubic and hexagonal lattices, and looked
for the most stable among them for several values of $\lambda$ between
$\lambda=1.1$ and $\lambda=3$.  Note that their calculations were
performed at fixed pressure $P$ rather than at fixed density, so that
the most stable structure corresponds to the minimum of the chemical
potential $\mu$. The minimum value of $\mu$ was always found to be
given either by the bcc001($\lambda$) or fcc001($\lambda$), with a
fcc-type to bcc-type transition on increasing the pressure. In their
Table~I, PS report the results obtained by minimizing with respect to
$\alpha$ and $\rho$ eleven different structures for $\lambda=3$ and
two different pressures, $P^{*}=0.05$ and $P^{*}=0.20$. According to
Eq.~(\ref{rhox}), for this value of $\lambda$ the fcc-bcc transition
density is $\rho_{c}^{*}(\lambda=3)=0.060$, and the transition
pressure obtained by differentiating $E/N$ with respect to $v=1/\rho$
is found to be $P_{c}^{*}(\lambda=3)=0.018$, which is lower than both
the values above. Hence, in both cases we expect the bcc001($\lambda$)
lattice to prevail, in agreement with PS results.  By minimizing $\mu$
with respect to $\rho$ at fixed $P$ for the bcc001($\lambda$) via the
relation $\mu=E/N+P/\rho$, we obtained $\mu^{*}=0.855718$ and
$\mu^{*}=2.093693$ for $P^{*}=0.05$ and $P^{*}=0.20$ respectively, in
very close agreement with PS results $\mu^{*}=0.855724$ and
$\mu^{*}=2.093695$.

\subsubsection{Degeneracy}
\label{subsubsec:PS_degeneracy}

Interestingly, PS also acknowledged the occurrence of degenerate
structures.  In fact, in their Table~I the bcc001($3$), bcc110($3$),
bcc111($3$), sc111($1.5$), and fcc001($3/\sqrt{2}$) are all
degenerate. The same is true for the sc001($3$) and the sc110($3$) and
for the fcc110($3$) and the fcc111($3$), even though the last two
groups do not correspond to the ground state.  PS pointed out that
{\em ``an emergent aspect of this table is the existence of a rich
  degeneracy that is only partly a result of the effective identity of
  crystal structures up to a dilation''} and that {\em ``points in
  these three lattices} [i.e., bcc001, bcc110, and bcc111] {\em have
  different local environments, as can be checked by counting the
  $n$th order neighbors for $n$ up to $4$, yet the three stretched
  crystals of minimum $\mu$ share the same $U/N$''} ($E/N$ in our
notation). They also commented {\em ``this fact is an emergent
  phenomenon whose deep reason remains unclear to us: it should deal
  with the dependence of $u$} (the potential $\Phi$ in our notation)
{\em on the ratio $r/\sigma(\vartheta)$, since the same symmetry holds
  with a polynomial, rather than Gaussian, dependence''}.

The origin of this degeneracy has been identified in
Sec.~\ref{subsec:degeneracy}.  Moreover, we are in a position to
explain not only why the degeneracy is there, but also why it involves
in particular the lattices found by PS, as we are discussing in detail below.

Let us first consider the structures obtained by deforming the bcc
lattice. As observed above, the bcc001($\lambda$) lattice is just that
of Eq.~(\ref{bccani}), which is obtained by identifying ${\bf A}_{0}$
with the matrix of the primitive vectors of the bcc lattice ${\bf
  A}_{0,{\rm bcc}}$, and deforming it along the nematic axis by a
factor $\lambda$.  In order to stretch the lattice vectors by a factor
$\alpha$ along a generic direction specified by a unit vector  
${\bf v}$ so as to obtain the bcc$\,
v_{x}v_{y}v_{z}$($\alpha$) structure, one has to multiply by $\alpha$
their components along ${\bf v}$, and leave unchanged those orthogonal
to ${\bf v}$. This amounts to applying to ${\bf A}_{0}$ the matrix
${\bf V}$ given by 
\begin{equation}               
{\bf V}=
\begin{pmatrix}
1+(\alpha-1)v_{x}^{2}   & (\alpha-1)v_{x}v_{y} & (\alpha-1)v_{x}v_{z} \\
(\alpha-1)v_{x}v_{y}   & 1+(\alpha-1)v_{y}^{2} & (\alpha-1)v_{y}v_{z} \\
(\alpha-1)v_{x}v_{z}   & (\alpha-1)v_{y}v_{z} & 1+(\alpha-1)v_{z}^{2} 
\end{pmatrix} 
\label{stretch}
\end{equation}
and adjusting the lattice constant $a$ in ${\bf A}_{0}$ so that at a
given density $\rho$ one has $v_{0}=\det{\bf A}_{0}=1/(\alpha\rho)$,
as discussed in Sec.~\ref{subsec:mapping}. Obviously, ${\bf V}$ is
diagonalized by rotating the axis vectors 
${\bf e}_{x}$, ${\bf e}_{y}$, ${\bf e}_{z}$ in
such a way that one of them, say ${\bf e}_{z}$, is mapped into ${\bf
  v}$, whereas ${\bf e}_{x}$ and ${\bf e}_{y}$ are mapped into two
orthogonal vectors ${\bf v}'$, ${\bf v''}$ lying in the plane
orthogonal to ${\bf v}$. We have then ${\bf D}_{\alpha}={\bf
  R}_{0}{\bf V}{\bf R}_{0}^{-1}$, where ${\bf D}_{\alpha}$ is the
diagonal matrix of Eq.~(\ref{diag}) with $\lambda$ replaced by
$\alpha$, and ${\bf R}_{0}$ is the rotation matrix connecting the old
and new axes such that
\begin{equation}
({\bf e}_{x}|{\bf e}_{y}|{\bf e}_{z})=({\bf v}|{\bf v}'|{\bf v}''){\bf R}_{0}  
\, .
\label{rot1}
\end{equation}
If ${\bf v}$ does not coincide with ${\bf e}_{z}$, in which case of
course ${\bf V}$ is already diagonal, ${\bf R}_{0}$ can be chosen as
\begin{equation}
{\bf R}_{0}=
\begin{pmatrix}
  \displaystyle{\frac{v_{y}}{\sqrt{1-v_{z}^{2}}}} & -
  \displaystyle{\frac{v_{x}}{\sqrt{1-v_{z}^{2}}}} 
& 0 \\
\displaystyle{\frac{v_{x}v_{z}}{\sqrt{1-v_{z}^{2}}}} \,  & 
\ \  \displaystyle{\frac{v_{y}v_{z}}{\sqrt{1-v_{z}^{2}}}} \, & -\sqrt{1-v_{z}^{2}} \\
v_{x} & v_{y} & v_{z}  
\end{pmatrix}
\, .
\label{rot2}
\end{equation}

After stretching the lattice, thus obtaining the matrix ${\bf V}{\bf
  A}_{0}$ of the stretched lattice vectors, the above described PS
recipe requires that the nematic axis be aligned with ${\bf v}$. In
the present notation, the nematic axis coincides with ${\bf e}_{z}$,
and this is aligned with ${\bf v}$ via the rotation~(\ref{rot1}). The
matrix of the primitive vectors ${\bf A}_{\rm PS}$ which identifies
the bcc$\, v_{x}v_{y}v_{z}$($\alpha$) lattice thus obtained is then
\begin{equation}
{\bf A}_{\rm PS}={\bf R}_{0}{\bf V}{\bf A}_{0}=
{\bf R}_{0}{\bf R}_{0}^{-1}{\bf D}_{\alpha}{\bf R}_{0}{\bf A}_{0}=
{\bf D}_{\alpha}{\bf R}_{0}{\bf A}_{0} \, .
\label{ps1}
\end{equation}
If ${\bf A}_{0}$ is expressed via Eq.~(\ref{A0}) by the matrix ${\bf
  A}$ (whose explicit expression is ${\bf A}_{{\rm bcc},\lambda}$ of
Eq.~(\ref{bccani}) for the specific case of the bcc lattice), one gets
\begin{equation}
{\bf A}_{\rm PS}=
{\bf D}_{\alpha}{\bf R}_{0}{\bf D}_{\lambda}^{-1}{\bf A} \, .
\label{ps2}
\end{equation}
For $\alpha=\lambda$, Eq.~(\ref{ps2}) is the same as
Eq.~(\ref{trasdir}) relating degenerate structures ${\bf A}$ and ${\bf
  A}'$ when the rotation matrix ${\bf R}$ is identified with ${\bf
  R}_{0}$. Therefore, the bcc001($\lambda$), bcc110($\lambda$), and
bcc111($\lambda$) are degenerate for arbitrary $\lambda$. By the same
token, this is still true if ${\bf A}_{0}$ is the matrix of a generic
Bravais lattice other than the bcc, irrespective of whether the
corresponding matrix ${\bf A}_{\rm PS}$, obtained by setting
$\alpha=\lambda$ in Eq.~(\ref{ps1}), corresponds to the ground state
or not. So, for instance, the sc001($\lambda$) and sc110($\lambda$)
are degenerate, as well the fcc(110)($\lambda$) and
fcc(111)($\lambda$), as pointed out above.  Besides, it is not even
necessary that ${\bf v}$ corresponds to some ``special'' direction of
high symmetry such as those considered by PS: according to
Eq.~(\ref{trasdir}), {\em any} direction goes, so that the degeneracy
is actually infinite, as discussed in Sec.~\ref{subsec:degeneracy}.

In addition, Table~I in Ref.~\cite{Prestipino:2007} shows that
deforming the sc or fcc lattices can again give the ground state, even
though in that case the stretching factor $\alpha$ does not coincide
with $\lambda$. As stated above, this happens for the sc111($1.5$) and
fcc001($3/\sqrt{2})$ lattices.  Such an occurrence can also be
explained by the present analysis.

The matrix ${\bf A}_{\rm PS}^{\rm sc111(\alpha)}$ of the
sc111($\alpha$) lattice is obtained from Eq.~(\ref{ps1}) when ${\bf
  A}_{0}$ is chosen as the unit cell ${\bf A}_{0,{\rm sc}}$ of the sc
lattice and ${\bf R}_{0}$ is the rotation matrix of Eq.~(\ref{rot2})
corresponding to the $111$ direction, i.e., to
$u_{x}=u_{y}=u_{z}=1/\sqrt{3}$. This gives
\begin{equation}
{\bf A}_{\rm PS}^{\rm sc111(\alpha)}=\frac{\ell_{\rm sc}}{\sqrt{3}}
\begin{pmatrix}
\displaystyle{\sqrt{\frac{3}{2}}} & -\displaystyle{\sqrt{\frac{3}{2}}} & \ 0 \\
\displaystyle{\frac{1}{\sqrt{2}}} & \ \ \  \displaystyle{\frac{1}{\sqrt{2}}} & -\sqrt{2} \\
\alpha & \ \ \alpha & \ \ \alpha 
\end{pmatrix} \, ,
\label{sc111}
\end{equation}
where the edge $\ell_{\rm sc}$ of the cell is given by
Eq.~(\ref{ellsc}) with $\lambda$ replaced by $\alpha$.  In order for
this lattice to have the same energy as the bcc001($\lambda$), a
sufficient condition is that the matrix ${\bf A}_{\rm PS}^{\rm
  sc111(\alpha)}$ and the matrix ${\bf A}_{{\rm bcc},\lambda}$ given
by Eq.~(\ref{bccani}) fulfill Eq.~(\ref{trasdir}). Using the relation
${\bf A}_{0,{\rm bcc}}={\bf D}_{\lambda}^{-1}{\bf A}_{{\rm
    bcc},\lambda}$, this amounts to requiring that there exists some
rotation matrix ${\bf R}$ such that
\begin{equation}
{\bf A}_{\rm PS}^{\rm sc111(\alpha)}={\bf D}_{\lambda}{\bf R}{\bf A}_{0,{\rm bcc}}  \, ,
\label{ps3}
\end{equation}
or, equivalently,
\begin{equation}
{\bf D}_{\lambda}^{-1}{\bf A}_{\rm PS}^{\rm sc111(\alpha)}{\bf A}_{0,{\rm bcc}}^{-1}={\bf R} \, ,
\label{ps4}
\end{equation}

It is readily checked that for $\alpha=\lambda/2$ Eq.~(\ref{ps4}) does
give a rotation matrix.

The fcc001($\alpha$) lattice is obtained from Eq.~(\ref{ps1}) by
choosing ${\bf A}_{0}$ as the unit cell ${\bf A}_{0,{\rm fcc}}$ of the
fcc lattice, and ${\bf R}_{0}$ as the identity.  One has then ${\bf
  A}_{\rm PS}^{\rm fcc001(\alpha)}={\bf A}_{{\rm fcc},\alpha}$, where
${\bf A}_{{\rm fcc},\alpha}$ is the same as ${\bf A}_{{\rm
    fcc},\lambda}$ given by Eq.~(\ref{fccani}), provided $\lambda$ is
replaced by $\alpha$.

Unlike ${\bf A}_{\rm PS}^{\rm sc111(\alpha)}$ with $\alpha=\lambda/2$,
${\bf A}_{{\rm fcc},\alpha}$ with $\alpha=\lambda/\sqrt{2}$ does {\em
  not} satisfy Eq.~(\ref{ps4}). However, it does satisfy its
generalized form Eq.~(\ref{trasdir3}), i.e., it can be checked that
for $\alpha=\lambda/\sqrt{2}$ there exists a rotation matrix ${\bf R}$
such that
\begin{equation} 
({\bf D}_{\lambda}{\bf R}{\bf A}_{0,{\rm bcc}})^{-1}{\bf A}_{{\rm fcc},\alpha}={\bf L} \, ,
\label{ps5}
\end{equation}
where ${\bf L}$ is a matrix with integer elements.

It should be noted that, unlike in the case of structures obtained by
deforming the same lattice along different directions such as the
bcc001($\lambda$), bcc110($\lambda$), and bcc111($\lambda$) discussed
above, when comparing structures obtained by deforming different
lattices, the direction of stretching ${\bf v}$ is indeed relevant to
the degeneracy: not any direction goes.  For instance, while the
sc111($\lambda/2$) and the bcc001($\lambda$) are degenerate, the
sc001($\lambda/2$) and the bcc001($\lambda$) are not.

In summary, our findings are fully consistent with those of PS at zero
temperature: the lattices which correspond to the ground state include
those singled out by PS, which are then found to be degenerate. At the
same time, the present analysis provides a rigorous proof of their
results, and shows that the degeneracy which they observed actually
involves infinitely many structures and, as they correctly argued, is
not peculiar to the Gaussian potential.

\subsection{Relation with previous investigations: Nikoubashman and Likos}
\label{subsec:nikoubashman}

In Ref.~\cite{Nikoubashman:2010}, NL investigated the ground state of
the aspherical Gaussian model using a genetic algorithm (GA) to
minimize the internal energy -- see Eq.~(\ref{lsum}) -- at fixed
density with respect to the lattice vectors and the direction ${\bf
  u}$ of the nematic axis. The minimization was performed for
densities in the interval $0<\rho^{*}<0.30$ and asphericity parameters
$\lambda$ in the interval $1<\lambda<3$. The main conclusions of their
study are conveyed by the phase diagram in the $(\rho, \lambda)$-plane
displayed in their Fig.~5(a), and can be summarized as follows:
\begin{itemize}
\item[(i)] GA minimization yields configurations of lower internal
  energy than those considered by PS. Therefore, PS configurations do
  not correspond to the ground state. This is displayed in Figs.~4 and
  14 of Ref.~\cite{Nikoubashman:2010}, where the difference between
  the internal energy per particle of PS bcc001($\lambda$) and
  fcc001($\lambda$) lattices and that obtained by GA is plotted as a
  function of $\rho$ for $\lambda=1.08$ and $\lambda=2.0$, and is
  indeed found to be always positive.
\item[(ii)] For every $\lambda$, a transition takes place which, in
  the limit $\lambda\rightarrow 1$, coincides with the aforementioned
  fcc-bcc transition of the spherical potential. The transition
  density $\rho_{c}(\lambda)$ is empirically found to be accurately
  represented by Eq.~(\ref{rhox}).
\item[(iii)] On each side of the $\rho_{c}(\lambda)$ curve, two
  distinct phases are present, which are identified by Roman numerals:
  phases I and III for $\rho<\rho_{c}(\lambda)$, and phases II and IV
  for $\rho>\rho_{c}(\lambda)$. The transition from phase I to phase
  III and that from phase II to phase IV take place on increasing
  $\lambda$, and in both cases the values of $\lambda$ corresponding
  to the transition are independent of $\rho$.
\item[(iv)] In addition, the bcc phase is found to survive in a small
  domain for $\rho>\rho_{c}(\lambda)$ and $\lambda$ close to $1$.
\end{itemize}

Clearly, point~(i) contradicts the findings of the present study, and
as such must be tackled.  The contradiction is all the more
surprising, in the light of the fact that point~(ii) is instead fully
consistent with our analysis, which would in fact provide an
explanation for it. In order to clarify this issue, we performed the
minimization of Eq.~(\ref{lsum}) for $\lambda=1.08$ and $\lambda=1.5$
in the interval $0<\rho^{*}<0.30$ using the GA code employed by NL. As
observed above, the value $\lambda=1.08$ is the same as that
considered in Fig.~4 of Ref.~\cite{Nikoubashman:2010}.  The results
were compared with those obtained here for the lattices generated by
the matrices ${\bf A}_{{\rm bcc},\lambda}$ and ${\bf A}_{{\rm
    fcc},\lambda}$ of Eqs.~(\ref{bccani}) and (\ref{fccani}) for
$\rho>\rho_{c}(\lambda)$ and $\rho<\rho_{c}(\lambda)$ respectively. We
found that at very low density, $\rho^{*}\lesssim 0.015$, the two
procedures gave the same energy within all the significant digits of
the GA results.  At larger density, the lattices corresponding to
${\bf A}_{{\rm bcc},\lambda}$ and ${\bf A}_{{\rm fcc},\lambda}$ always
gave the lower energy, but even in that case the difference was found
to be at most of the order of $\sim 10^{-8}$.  This is of the same
order of magnitude as the difference obtained for $\lambda=1$, i.e.,
the spherical Gaussian potential, between the energy predicted by the
GA and that of the bcc or fcc lattices, which are known to give the
exact ground state in the spherical case. Therefore, we conclude that
the lattice energies predicted by the GA are actually the same as
those of the lattices generated by ${\bf A}_{{\rm bcc},\lambda}$ and
${\bf A}_{{\rm fcc},\lambda}$, and that the tiny discrepancy is due to
small numerical errors of the GA, which are present also in the
spherical case, possibly due to roundoff. Since, as already discussed
in Sec.~\ref{subsec:prestipino}, these lattices coincide with PS
bcc001($\lambda$) and fcc001($\lambda$) lattices, we must similarly
conclude that NL energies are the same as those of PS
bcc001($\lambda$) or fcc001($\lambda)$. This conclusion cannot be
reconciled with NL point~(i), since the difference between the
internal energy per particle of PS lattices and that of NL lattices
plotted in their Figs.~4 and 14 is several order of magnitues larger
than the extremely small discrepancy reported above, and cannot be
traced back to numerical accuracy. Therefore, the only possible
explanation left to us, is that there must have been some error
in the calculation of the internal energies
of PS lattices as performed by NL, which prevented them 
from acknowledging that they actually
coincide with those obtained by GA minimization.

If the configurations found by NL and those found here are indeed
degenerate, then for our picture to be correct NL lattices must be
obtained from ours by the transformation described in
Sec.~\ref{subsec:degeneracy}. This means that the corresponding
matrices have to be related by Eq.~(\ref{trasdir3}). In order to
perform this check, a preliminary observation must be made: whereas in
the present analysis the axes are chosen in such a way that the
nematic direction ${\bf u}$ coincides with the $z$-axis, in NL
calculation they were instead chosen so that the direction of one of
the primitive vectors coincides with the $x$-axis.  Before feeding NL
primitive vectors as given by their GA minimization into
Eq.~(\ref{trasdir3}), we then have to express them on our basis by
replacing the original matrix ${\bf A}_{\rm NL}$ of NL primitive
vectors with ${\bf R}_{\rm NL}{\bf A}_{\rm NL}$, where ${\bf R}_{\rm
  NL}$ is the rotation which alignes NL $z$-axis with ${\bf u}$. If
$\alpha$ and $\beta$ are the polar and azimuthal angles of ${\bf u}$
with respect to NL axes, ${\bf R}_{\rm NL}$ is given by
\begin{equation}  
{\bf R}_{\rm NL}=
\begin{pmatrix}
\ \cos\alpha\cos\beta \ & \cos\alpha\sin\beta \ & -\sin\alpha \\
\ -\sin\beta & \ \cos\beta & \ 0  \\
\ \sin\alpha\cos\beta & \ \sin\alpha\sin\beta & \ \cos\alpha 
\end{pmatrix} \, .
\label{rotnl}
\end{equation}
Eq.~(\ref{trasdir3}) then becomes
\begin{equation} 
({\bf D}_{\lambda}{\bf R}{\bf A}_{0,{\rm bcc}})^{-1}{\bf R}_{\rm NL}{\bf A}_{\rm NL}={\bf L} \, ,
\label{trasdirnl}
\end{equation}
or the similar one with ${\bf A}_{0,{\rm bcc}}$ replaced by ${\bf
  A}_{0,{\rm fcc}}$ depending on whether one has
$\rho>\rho_{c}(\lambda)$ or $\rho<\rho_{c}(\lambda)$, where ${\bf
  A}_{0,{\rm bcc}}$ and ${\bf A}_{0,{\rm fcc}}$ are as usual the
standard matrices of the bcc and fcc lattices, in which the edge of
the conventional cubic cell is given by Eqs.~(\ref{ellbcc}) and
(\ref{ellfcc}), and ${\bf A}_{\rm NL}$, ${\bf R}_{\rm NL}$ are
obtained from the output of the GA minimization.

We did verify that Eq.~(\ref{trasdirnl}) is always fulfilled, i.e., it
was always possible to find some rotation matrix ${\bf R}$ such that
${\bf L}$ would consist of integer elements.  This provides compelling
evidence that the lattices determined by NL conform to the general
picture described in Sec.~\ref{subsec:degeneracy}.  We can now examine
the above points (iii) and (iv) in the light of this conclusion.

As for point (iii), the implication is that phases~I and III are
necessarily degenerate, so that there is not an actual transition
between them: both phases are found in the whole region
$\rho<\rho_{c}(\lambda)$, and besides them infinitely many other
degenerate phases exist. The same applies to phases~II and IV for
$\rho>\rho_{c}(\lambda)$. We surmise that, while GA always provides a
genuine ground state of the system, nothing prevents it from
``jumping'' from a ground state to another as $\lambda$ is
changed. But these jumps are just a consequence of the degeneracy, and
do not represent a true phase transition, whereby a state would become
favored with respect the other.

\begin{figure}
\includegraphics[width=10cm]{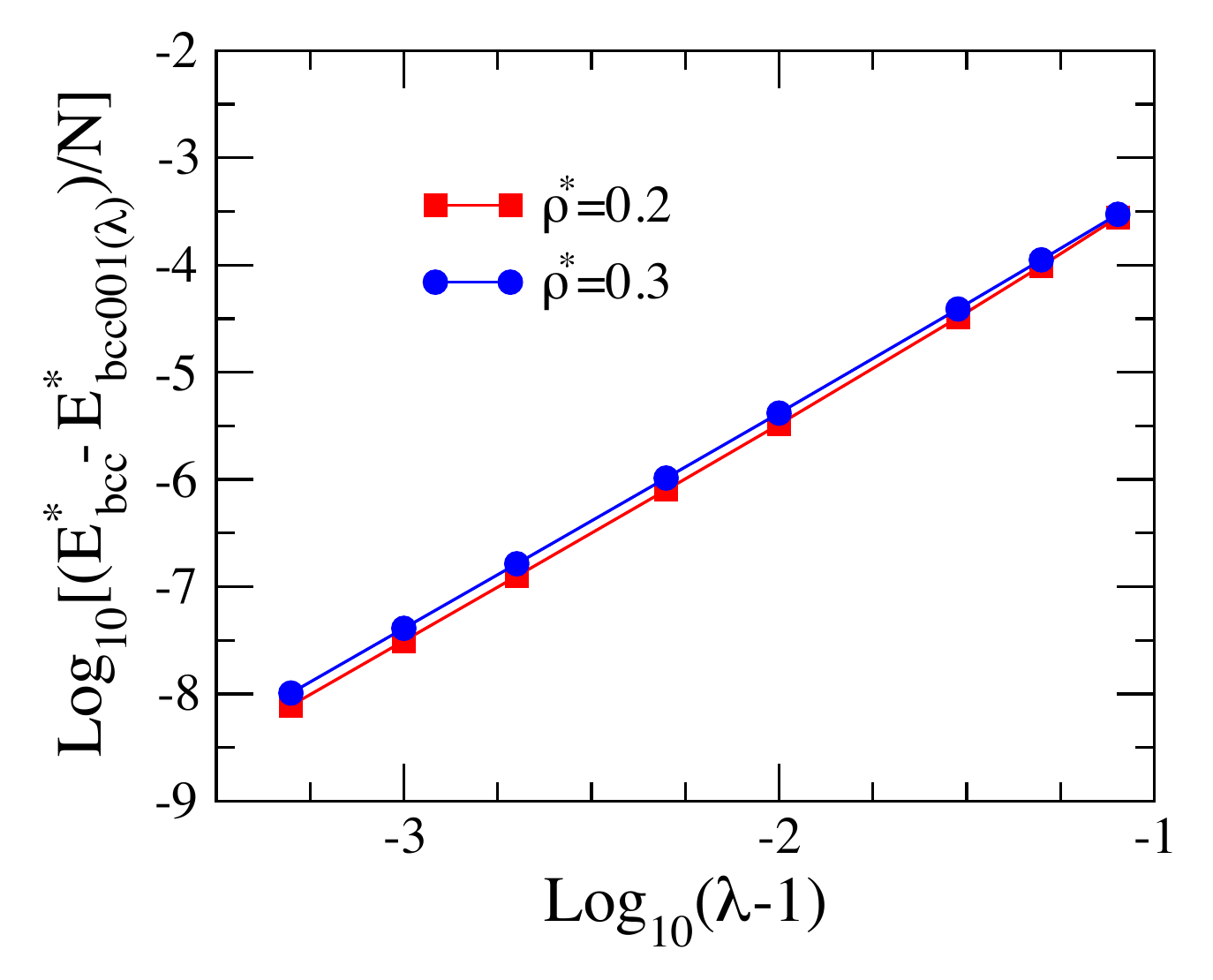}
\caption{Log-log plot of the difference between the internal energy
  per particle of the orientation-optimized bcc lattice and that of
  the bcc001($\lambda$) lattice of the aspherical Gaussian model for
  $\lambda$ close to $1$ and $\rho^{*}=0.2$, $\rho^{*}=0.3$.}
\label{fig:cfr_energy}
\end{figure}

Point (iv) is also at odds with the present picture, according to
which the bcc001($\lambda$) must prevail over the non-deformed bcc
lattice in the whole region $\rho>\rho_{c}(\lambda)$. This is indeed
what happens, as displayed in Fig.~\ref{fig:cfr_energy}, where the
difference between the internal energy per particle of the bcc lattice
optimized with respect to the direction ${\bf u}$ of the nematic axis
and that of the bcc001($\lambda$) lattice has been plotted as a
function of $\lambda$ for $\lambda$ close to $1$ at two densities,
both of which should lie in the ``pocket'' of stability of the bcc
according to NL.  At variance with NL result, the bcc001($\lambda)$ is
always found to give the lower energy, even for very small degree of
anisotropy. At the same time, the figure shows that in this limit the
energy difference becomes extremely small, and we deem it likely that
it simply goes beyond the numerical accuracy of the GA, thereby
leading to the incorrect identification of the bcc as the more stable
structure.

We may then say that NL results need to be reconsidered in two
respects: first, the lattices obtained by their GA-based approach do
not have a lower energy than that of the lattices singled out by PS as
the best candidates for the ground state: their energy is actually the
same as that of PS lattices, and both NL and PS lattices represent
true ground states of the model; second, the infinite degeneracy of
those structures should be acknowledged. At the same time, the ability
of GA minimization to converge on (one of) the lowest-energy
structure(s) in a completely unbiased way, is one more example of the
predictive power of these algorithms.

\section{Conclusions}
\label{sec:conclusions}

We have studied the ordered configurations at vanishing temperature of
a system of Gaussian core nematics (GCN), where ellipsoids of
revolution (spheroids) with aspect ratio $\lambda$ interact via a
generalized Gaussian potential, such that the directional unit vectors
${\bf u}$ of the particles are oriented in a mutually parallel manner.
We showed that, by rescaling the lattice vectors along the direction
of ${\bf u}$, the crystal phases of this system of aspherical
particles are mapped into those of the system of spherical particles
with $\lambda=1$. In the mapping the energy is conserved, whereas the
densities of the systems of aspherical and spherical particles, $\rho$
and $\rho_0$, are related via $\rho_0=\lambda \rho$. Since the ordered
configurations corresponding to the ground state of the Gaussian fluid
of spherical particles are known exactly, this enables one to obtain
straightforwardly the ground state of the GCN.

Based on this mapping and the ensuing conclusions we could furthermore
demonstrate that the ordered configurations of the GCN must be
infinitely degenerate, i.e., there is an infinite number of Bravais
lattices, characterized by the same energy. This multitude of
configurations is obtained in the following manner: starting from a
given Bravais lattice of the {\it aspherical} potential one constructs
the related Bravais lattice of the {\it spherical} potential by
rescaling the lattice vectors; in a subsequent step, this lattice is
rotated via an arbitrary rotation matrix, leading to a Bravais lattice
of the same kind as the original one. Transforming these lattice
vectors back leads to a new ordered phase of the aspherical
particles. As all these steps preserve the internal energy, the
resulting ordered phases must be infinitely degenerate. The mechanism
leading to the degeneracy was in fact already pointed out by Schiller
{\it et al.}  in their investigations of the crystal phases of
hard-core spheroids with contact
interactions~\cite{Schiller:2011,Schiller:2012}, although its
relevance to the GCN has not, to our knowledge, been pointed out
before.

With the help of these results we could elucidate and fully explain
the findings of two previous contributions dedicated to the GCN by
Prestipino and Saija (PS) \cite{Prestipino:2007} and Nikoubashman and
Likos (NL) \cite{Nikoubashman:2010}.  In particular, we proved that
the lattices corresponding to the ground state do include the
configurations singled out by PS as their best candidates, explained
the origin of the degeneracy which they observed, and showed that it
must be actually infinite.  Moreover, we could explain the occurrence
of the phase transition found by NL on changing the density, as well
as the dependence of the transition density on $\lambda$ empirically
found by them.  Further, by re-analysing the data of these studies we
found that some alleged inconsistency between the ground state
configurations identified in the two papers is actually not there: the
energy of those configurations is rigorously the same, and all of
them are obtained by rescaling according to the procedure outlined
above either the fcc or the bcc lattice, which provide the ground
state of the spherically symmetric Gaussian potential at low and high
density respectively.  The apparent discrepancy must be traced back to
some glitch in the calculation of the energy of PS configurations as
reported by NL, possibly due to issues of numerical accuracy.

The analysis developed here for ellipsoids of revolution can be
straightforwardly extended to aspherical Gaussian particles whose
shape is described by a generic ellipsoid, provided the orientation of
the ellipsoids remains the same for all particles, as in the present
model. Moreover, it is not peculiar to the Gaussian potential, and
could be equally well applied to different soft-core interactions,
including the cluster-forming potentials of the so-called $Q^{\pm}$
class~\cite{Likos:2001} such as the generalized exponential model of
order four (GEM-4) considered in Ref.~\cite{Weissenhofer:2018}, even
though the resulting phase diagram will of course be different from
that described in this contribution.

The assumption that all particles share the same orientation from the
outset is clearly a serious limit of the GCN model. Nevertheless, its
phase behavior and the related degeneracy may still have some bearing
on that of more realistic models, {\it if} nematic ordering is indeed
preferred in the crystal phases.  In their aforementioned study of
crystals of parallel hard-core
spheroids~\cite{Schiller:2011,Schiller:2012}, Schiller {\it et al.}
observed that the existence of an infinite number of different
lattices with the same energy can be a source of disorder, hindering
the formation of crystals with a specified structure, and their
observation is equally relevant to the soft-core model considered
here.  At the same time, such a feature could also have interesting
consequences on the vibrational, elastic, and optical properties of
those systems.

A question which comes naturally is, whether this infinite degeneracy
may survive even at non-vanishing temperature $T$. For the GCN, the
simulation results obtained by PS~\cite{Prestipino:2007} indicate that
the degeneracy is lifted as soon as one has $T>0$. However, it is
possible that in this respect the specific form of the interaction
may play some role, as seems to be the case, for instance, for the
aspherical GEM-4 particles of Ref.~\cite{Weissenhofer:2018}. We plan
to come back to this issue in the near future.

\section*{Acknowledgments}

GK acknowledges financial support by the Austrian Science Foundation,
FWF, under Proj.No. I3846.  DP acknowledges financial support by
Universit\`a degli Studi di Milano under Project
PSR2019\rule{0.15cm}{0.4pt}DIP\!\rule{0.15cm}{0.4pt}008-Linea~2.  The
authors would like to thank Arash Nikoubashman (Mainz) for making
available the code which represents the basis of the data presented in
Ref.~\cite{Nikoubashman:2010} and Alberto Parola (Como Insubria) for
useful conversation and his critical reading of the preliminary notes
of the manuscript.  DP wishes to thank Paola Dotti for her interest
and encouragement. Before submission the manuscript has been
distributed to Santi Preistipino and Franz Saija (both Palermo), as
well as to Arash Nikoubashman (Mainz) and Christos N. Likos (Vienna):
we gratefully acknowledge the constructive remarks raised by these
colleagues.

\end{document}